\documentclass[prl,aps,twocolumn,superscriptaddress,showpacs]{revtex4}
\usepackage{amsmath,amssymb,graphicx}
\usepackage{pxfonts}
\usepackage{graphicx}
\usepackage{color}
\usepackage{float}

\begin{document}

\date{\today}

\title{Attractive and repulsive dipolar interaction in bilayers of indirect excitons}

\author{D.\,J.~Choksy}
\affiliation{Department of Physics, University of California at San Diego, La Jolla, California 92093-0319, USA}
\author{Chao~Xu}
\affiliation{Department of Physics, University of California at San Diego, La Jolla, California 92093-0319, USA}
\author{M.\,M.~Fogler}
\affiliation{Department of Physics, University of California at San Diego, La Jolla, California 92093-0319, USA}
\author{L.\,V.~Butov}
\affiliation{Department of Physics, University of California at San Diego, La Jolla, California 92093-0319, USA}
\author{J.~Norman}
\affiliation{Materials Department, University of California at Santa Barbara, Santa Barbara, California 93106-5050, USA}
\author{A.\,C.~Gossard}
\affiliation{Materials Department, University of California at Santa Barbara, Santa Barbara, California 93106-5050, USA}

\date{\today}

\begin{abstract}
\noindent We explore attractive dipolar interaction in indirect excitons (IXs). For one layer of IXs in a single pair of coupled quantum wells (CQW), the out-of-plane IX electric dipoles lead to repulsive dipolar interaction between IXs. The attractive dipolar interaction between IXs is realized in a 2-CQW heterostructure with two IX layers in two separated CQW pairs. We found both in experimental measurements and theoretical simulations that increasing density of IXs in one layer causes a monotonic energy reduction for IXs in the other layer. We also found an in-plane shift of a cloud of IXs in one layer towards a cloud of IXs in the other layer. This behaviour is qualitatively consistent with attractive dipolar interaction. The measured IX energy reduction and IX cloud shift are higher than the values given by the correlated liquid theory. 
\end{abstract}
\maketitle

A spatially indirect excitons (IX), also known as an interlayer exciton, is a bound pair of an electron and a hole confined in separated layers. Due to the electron-hole separation, IXs have built-in electric dipole moment $ed$, where $d$ is the distance between the electron and hole layers and $e$ electron charge. Furthermore, due to the electron-hole separation, IXs have long lifetimes within which they can cool below the temperature of quantum degeneracy~\cite{Lozovik1976}. These properties make IXs a platform for exploring quantum gases with dipolar interaction.

IXs can be realized in a pair of quantum wells separated by a narrow tunneling barrier. For one layer of IXs in a single pair of coupled quantum wells (CQW), the out-of-plane IX electric dipoles lead to repulsive dipolar interaction between side-to-side IX dipoles (Fig.~1). This configuration is extensively studied both theoretically and experimentally. The phenomena originating from the repulsive dipolar interaction in a single IX layer include the enhancement of IX energy with density that has been known since early studies of IXs~\cite{Yoshioka1990, Butov1994, Zhu1995, Lozovik1997, Butov1999}, screening of in-plane disorder potential by repulsively interacting IXs~\cite{Butov2002, Ivanov2002, Ivanov2006, Remeika2009} that leads to IX delocalization and long-range IX transport~\cite{Hagn1995, Butov2002, Ivanov2002, Voros2005, Ivanov2006, Gartner2006, Remeika2009, Lasic2010, Gorbunov2011, Alloing2012, Lasic2014, Dorow2018a}, strong correlations \cite{Ben-Tabou2003, Zimmermann2007, Schindler2008, Lozovik2007, Remeika2009, Laikhtman2009, Ivanov2010, Cohen2011, Remeika2015}, and predicted crystal phases~\cite{Lozovik1996, Palo2002, Astrakharchik2007, Schleede2012, Suris2016, d}. 

\begin{figure}
\begin{center}
\includegraphics[width=8.5cm]{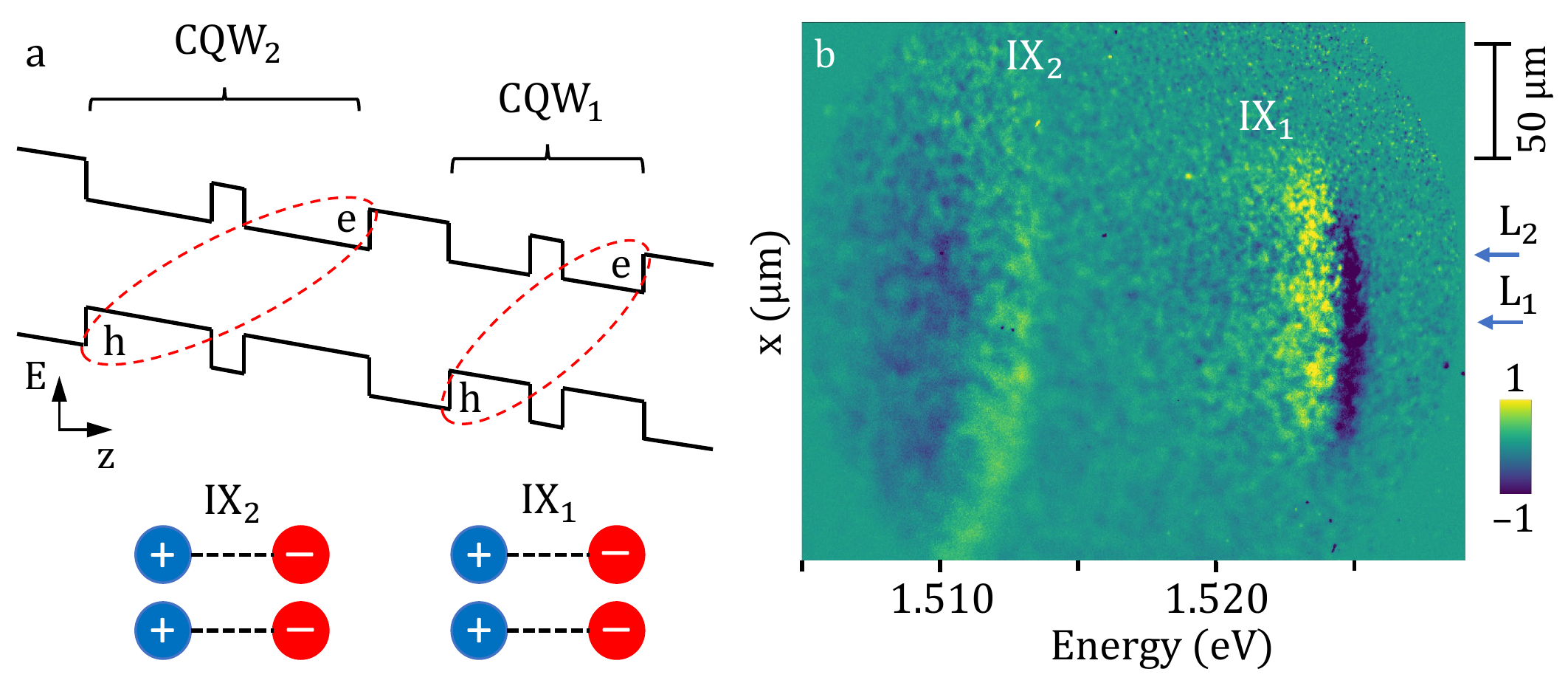}
\caption{\textbf{2-CQW heterostructure and $x$-energy image.} 
(a) Diagram of 2-CQW structure with two CQW pairs. IX$_2$ form in CQW$_2$, IX$_1$ in CQW$_1$. 
Schematic below shows IX dipoles. $-$ electrons, $+$ holes. The intra-CQW interaction between IX$_2$ (or IX$_1$) side-to-side dipoles is repulsive. The inter-CQW interaction between IX$_2$ and IX$_1$ head-to-tail dipoles is attractive.
(b) Differential $x$-energy luminescence image. The arrows indicate the excitation spot positions of L$_2$ and L$_1$ lasers resonant to direct excitons in 15~nm CQW$_2$ and 12~nm CQW$_1$, respectively. L$_2$ generates IX$_2$. L$_1$ generates IX$_1$ and also a smaller concentration of IX$_2$. The laser powers $P_{\rm L1} = 10$~$\mu$W, $P_{\rm L2} = 250$~$\mu$W. The differential $x$-energy image is obtained by subtracting the $x$-energy images created by only L$_1$ on and by only L$_2$ on from the $x$-energy image created by both lasers on. The differential $x$-energy image shows an increase of IX$_2$ energy, a decrease of IX$_1$ energy, and a spatial shift of the IX$_1$ cloud towards the IX$_2$ cloud.
}
\end{center}
\label{fig:spectra}
\end{figure}

\begin{figure*}
\begin{center}
\includegraphics[width=15.5cm]{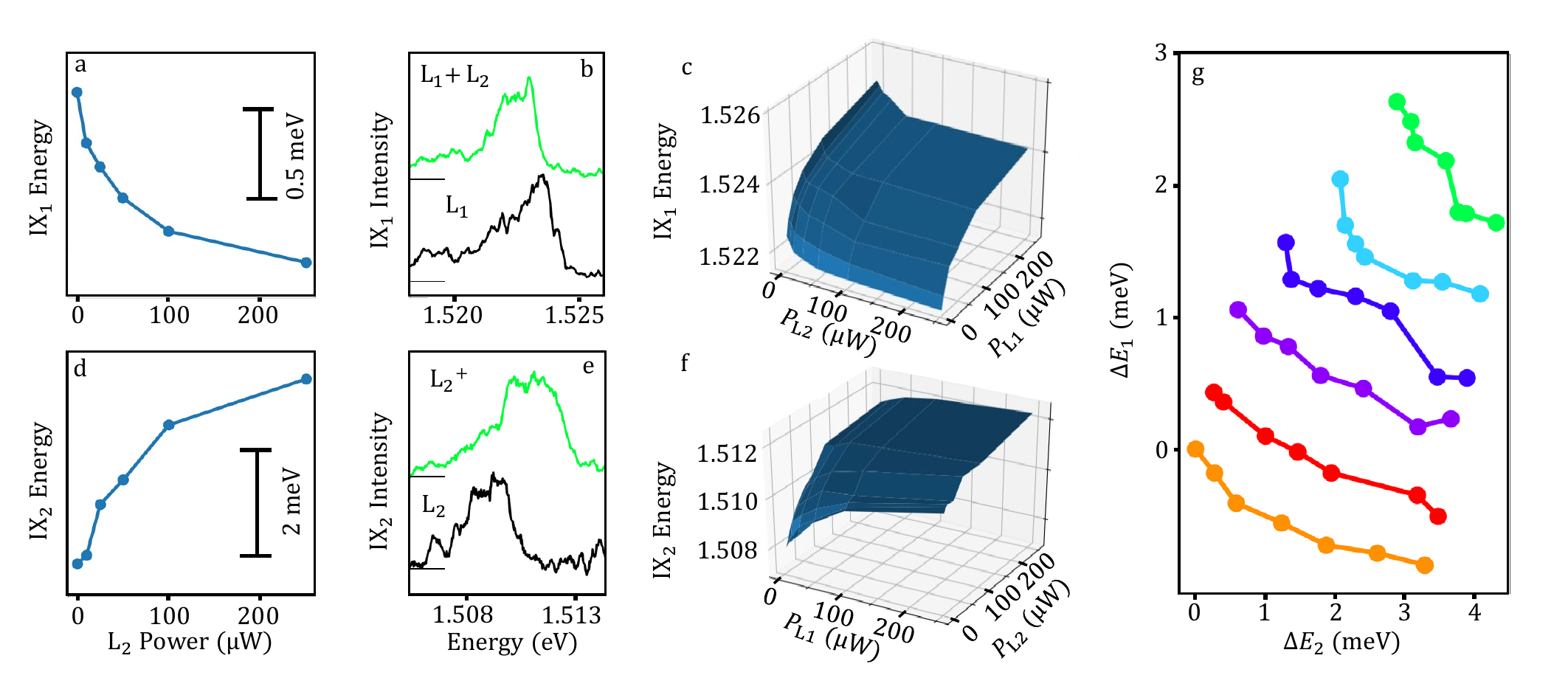}
\caption{\textbf{Decrease and increase of IX energy due to attractive and repulsive dipolar IX interaction: Experiment.} 
(a) The decrease of IX$_1$ energy with increasing L$_2$ power and, in turn, IX$_2$ density. $P_{\rm L1} = 10$~$\mu$W. An energy decrease corresponds to attractive ${\rm IX}_1 - {\rm IX}_2$ interaction. 
(b) The red shift of IX$_1$ spectrum with turning on L$_2$, which increases IX$_2$ density. Black line shows IX$_1$ spectrum when only L$_1$ is on, $P_{\rm L1} = 10$~$\mu$W. Green line shows IX$_1$ spectrum when additional L$_2$ is on, $P_{\rm L2} = 250$~$\mu$W. 
(c) IX$_1$ energy as a function of both $P_{\rm L1}$ and $P_{\rm L2}$.
(d) The increase of IX$_2$ energy with increasing L$_2$ power and, in turn, IX$_2$ density. $P_{\rm L1} = 0$. An energy increase corresponds to repulsive ${\rm IX}_2 - {\rm IX}_2$ interaction. 
(e) The blue shift of IX$_2$ spectrum with increasing L$_2$ power, which increases IX$_2$ density. $P_{\rm L2} = 25$~$\mu$W (black line) and 250~$\mu$W (green line). 
(f) IX$_2$ energy as a function of both $P_{\rm L1}$ and $P_{\rm L2}$. 
(g) The change in IX$_1$ energy vs the change of IX$_2$ energy. Each set of data corresponds to increasing L$_2$ power. For the sets of data presented by orange, red, purple, blue, cyan, and green points, L$_1 = 5, 10, 25, 50, 100,$ and 250~$\mu$W, respectively.
}
\end{center}
\label{fig:spectra}
\end{figure*}

Quantum gases with dipolar interaction are also explored with cold atoms. In these systems, dipolar interactions lead to droplet structures with spatial ordering and coherence~\cite{Kadau2016, Ferrier-Barbut2016, Tanzi2019, Bottcher2019, Chomaz2019}, few-body complexes~\cite{Wunsch2011, Dalmonte2011}, and pair superfluid and crystal phases in bilayers of dipoles~\cite{Lu2008, Safavi-Naini2013, Macia2014, Cinti2017}.

The specific property of dipolar interaction is its anisotropy. For instance, for two parallel dipoles tilted at angle $\theta$ relative to the line connecting them, the interaction at $r \gg d_1, d_2$ is given by $v(r) \sim e^2 d_1 d_2 (1 - 3 \cos^2{\theta}) / \epsilon r^3$, where $\epsilon$ is the dielectric constant of the material and $p_{1,2} = e d_{1,2}$ the dipole moments~\cite{3D}. For the out-of-plane IX dipoles in a single IX layer this expression reduces to $v(r) \sim e^2 d^2 / \epsilon r^3$ describing the repulsive dipolar interaction between IXs. 

The other specific property of dipolar interaction for the IXs is the induced orientation of IX dipoles. The heterostructure design and/or applied voltage, which produces the electric field in the heterostructure, determine the quantum well layers where electrons and holes are confined: Exchanging the quantum wells by the electron and the hole, i.e. flipping the IX dipole, is energetically unfavorable. Furthermore, tilting the IX dipole relative to the $z$ direction causes an in-plane separation of the electron and the hole in the IX and, as a result, reduces the IX binding energy. This induces the orientation of IX dipoles in the direction normal to the QW plane. 

The induced orientation of the IX dipoles and the repulsive dipolar interaction for a single IX layer make challenging exploring the attractive dipolar interaction in IX systems. The studies of IX dipoles have been concentrated on the case of repulsively interacting IXs~\cite{Lozovik1976, Yoshioka1990, Butov1994, Zhu1995, Lozovik1997, Butov1999, Hagn1995, Butov2002, Ivanov2002, Voros2005, Ivanov2006, Gartner2006, Lasic2010, Gorbunov2011, Alloing2012, Lasic2014, Dorow2018a, Ben-Tabou2003, Zimmermann2007, Schindler2008, Lozovik2007, Remeika2009, Laikhtman2009, Ivanov2010, Cohen2011, Remeika2015, Lozovik1996, Palo2002, Astrakharchik2007, Schleede2012, Suris2016}. However, the angle-dependent dipolar IX interaction and, in particular, dipolar attraction gives an access to new phenomena in quantum dipolar gases. For instance, the dipolar attraction leads to the phenomena in cold atoms outlined above~\cite{Kadau2016, Ferrier-Barbut2016, Tanzi2019, Bottcher2019, Chomaz2019, Wunsch2011, Dalmonte2011, Lu2008, Safavi-Naini2013, Macia2014, Cinti2017}. IX attraction can be realized by extending IX heterostructures beyond a single CQW design and studies of attractively interacting IX dipoles were recently started in two stacked CQW pairs~\cite{Cohen2016, Hubert2019, Hubert2020}.

In this work, we explore the attractive dipolar interaction between IXs in a 2-CQW heterostructure with two IX layers in two separated CQW pairs (Fig.~1). The intra-CQW interaction between IX side-to-side dipoles is repulsive similar to single CQW heterostructures~\cite{Lozovik1976, Yoshioka1990, Butov1994, Zhu1995, Lozovik1997, Butov1999, Hagn1995, Butov2002, Ivanov2002, Voros2005, Ivanov2006, Gartner2006, Lasic2010, Gorbunov2011, Alloing2012, Lasic2014, Dorow2018a, Ben-Tabou2003, Zimmermann2007, Schindler2008, Lozovik2007, Remeika2009, Laikhtman2009, Ivanov2010, Cohen2011, Remeika2015, Lozovik1996, Palo2002, Astrakharchik2007, Schleede2012, Suris2016}. The inter-CQW interaction between IX head-to-tail dipoles is attractive. It changes to repulsive with increasing in-plane separation between the IXs and, in turn, $\theta$ following the anisotropy of dipolar interaction outlined above. Both our experimental measurements and theoretical simulations show (i) a monotonic energy reduction for IXs in one layer with increasing density of IXs in the other layer and (ii) an in-plane shift of a cloud of IXs in one layer towards a cloud of IXs in the other layer. This behaviour is qualitatively consistent with attractive dipolar interaction, however, the measured IX energy reduction and IX cloud shift are higher than the values given by the correlated liquid theory. 

{\bf Experiment.} The studied 2-CQW heterostructure (Fig.~1a) is grown by molecular beam epitaxy. Indirect excitons IX$_2$ form in CQW$_2$, indirect excitons IX$_1$ in CQW$_1$. CQW$_2$ consist of two 15~nm GaAs QWs separated by 4~nm Al$_{0.33}$Ga$_{0.67}$As barrier, CQW$_1$ consists of two 12~nm GaAs QWs separated by 4~nm Al$_{0.33}$Ga$_{0.67}$As barrier. CQW$_2$ and CQW$_1$ are separated by 12~nm Al$_{0.33}$Ga$_{0.67}$As barrier, narrow enough to allow substantial inter-layer interaction between IX$_2$ and IX$_1$, yet wide enough to suppress tunneling of electrons and holes between CQW$_2$ and CQW$_1$. $n^+$-GaAs layer with $n_{\rm Si} \sim 10^{18}$~cm$^{-3}$ serves as a bottom electrode. The CQW pair is positioned 100~nm above the $n^+$-GaAs layer within undoped 1~$\mu$m thick Al$_{0.33}$Ga$_{0.67}$As layer. The two CQW pairs are positioned closer to the homogeneous bottom electrode to suppress the fringing in-plane electric field in excitonic devices~\cite{Hammack2006}. The top semitransparent electrode is fabricated by applying 2~nm Ti and 7~nm Pt on 7.5~nm GaAs cap layer. Applied gate voltage $V_{\rm g} = - 2$~V creates electric field in the $z$-direction. 

The IX$_2$ energy is lower than the IX$_1$ energy. This energy difference gives an opportunity to selectively generate IX$_2$ by optical excitation. Excitons are generated by semiconductor lasers L$_2$ and L$_1$ at the energies 1.532 and 1.541 eV resonant to spatially direct excitons (DXs) in CQW$_2$ and CQW$_1$, respectively. The resonant to DX excitation increases the light absorption and, in turn, IX density for a given laser power~\cite{Kuznetsova2012}. L$_2$ generates IX$_2$. L$_1$ generates IX$_1$ and also roughly 2 times smaller concentration of IX$_2$ due to a weaker nonresonant absorption of L$_1$ light in CQW$_2$. L$_2$ and L$_1$ excitations are focused to $\sim 5$~$\mu$m hwhm spots, which are separated by 50~$\mu$m. This configuration allows exploring the effects of IX interactions on the IX cloud position. IX photoluminescence (PL) is measured in a 20~ns time window starting 20~ns after the end of the L$_1$ and/or L$_2$ excitation pulses. This allows for studying of only long-lived IXs after DXs recombined. Both IX$_2$ and IX$_1$ have long lifetime in the range of hundreds of ns ($\sim 800$~ns for IX$_2$ and $\sim 260$~ns for IX$_1$) allowing them to travel over long distances reaching hundreds of microns. 

Time-resolved imaging experiments are performed with a laser pulse duration 2000~ns, period 4000~ns, and edge sharpness $\sim 2$~ns. The rectangular-shaped pulses are realized by a pulse generator driving the semiconductor lasers. The pulse duration and period are optimized to allow the IX PL image to approach equilibrium during the laser excitation and decay between laser pulses. The PL images are captured using a PicoStar HR TauTec time-gated intensifier. The PL passes through a spectrometer with a resolution of 0.18 meV before entering the intensifier coupled to a liquid-nitrogen-cooled CCD. The measurements are performed at $T_{\rm bath} = 1.7$~K.

To analyze the attractive inter-layer IX interaction in the IX bilayer we measure how the selective generation of IX$_2$ affects the energies and cloud position of IX$_1$. Figure~1b presents the differential $x$-energy image obtained by subtracting the $x$-energy images created by only L$_1$ on [Fig.~S5c in supporting information (SI)] and by only L$_2$ on (Fig.~S5b) from the $x$-energy image created by both lasers on simultaneously (Fig.~S5a). The differential $x$-energy image shows an increase of IX$_2$ energy, a decrease of IX$_1$ energy, and a spatial shift of the IX$_1$ cloud towards the IX$_2$ cloud. These phenomena are detailed below.

First, we consider the IX energy variations. Figures~2a-c show that the increase of L$_2$ power ($P_{\rm L2}$) and, in turn, IX$_2$ density ($n_2$) leads to a monotonic decrease of IX$_1$ energy. An energy decrease corresponds to attractive ${\rm IX}_1 - {\rm IX}_2$ interaction. In comparison, when only IX$_2$ are present in the system (L$_1$ is off), the increase of $P_{\rm L2}$ and, in turn, $n_2$ leads to a monotonic increase of IX$_2$ energy (Fig.~2d-f). An energy increase corresponds to repulsive ${\rm IX}_2 - {\rm IX}_2$ interaction, which has been extensively studied in single layers of IXs~\cite{Lozovik1976, Yoshioka1990, Butov1994, Zhu1995, Lozovik1997, Butov1999, Hagn1995, Butov2002, Ivanov2002, Voros2005, Ivanov2006, Gartner2006, Lasic2010, Gorbunov2011, Alloing2012, Lasic2014, Dorow2018a, Ben-Tabou2003, Zimmermann2007, Schindler2008, Lozovik2007, Remeika2009, Laikhtman2009, Ivanov2010, Cohen2011, Remeika2015, Lozovik1996, Palo2002, Astrakharchik2007, Schleede2012, Suris2016}. 

Figures~2c,f also show that the increase of $P_{\rm L1}$ leads to a monotonic increase of both IX$_1$ and IX$_2$ energies. $L_1$ generates both IX$_1$ and IX$_2$, therefore, the effect of increasing $P_{\rm L1}$ on IX$_1$ (or IX$_2$) energy is a combined effect of attractive ${\rm IX}_1 - {\rm IX}_2$ and repulsive ${\rm IX}_1 - {\rm IX}_1$ (or ${\rm IX}_2 - {\rm IX}_2$) interactions. The monotonic increase of both IX$_1$ and IX$_2$ energies with $P_{\rm L1}$ indicates that the repulsive interaction is stronger. This is consistent with the relative strength of the attractive (Fig.~2a) and repulsive (Fig.~2d) interaction in the experiments with increasing $P_{\rm L2}$ which increase only IX$_2$ density.

In the mean-field approximation the repulsive interaction between IXs in a single layer increases the IX energy by $\Delta E = 4 \pi e^2 d n / \epsilon$. This equation known as the “plate capacitor” formula provides a qualitative explanation for the observed monotonic increase of $\Delta E$ with the exciton density $n$~\cite{Yoshioka1990}. However, the capacitor formula can significantly overestimate $\Delta E(n)$ due to the IX correlations~\cite{Ben-Tabou2003, Zimmermann2007, Schindler2008, Lozovik2007, Remeika2009, Laikhtman2009, Ivanov2010, Cohen2011, Remeika2015}. To compare the attractive and repulsive dipolar interactions, avoiding the complexity of the relation between $\Delta E$ and $n$, Fig.~2g presents the change in IX$_1$ energy $\Delta E_1$ vs the change of IX$_2$ energy $\Delta E_2$ for the data in Fig.~2c,f. The energy shifts $\Delta E$ are measured relative to the IX energies at the lowest $n$. Figure~2g shows that for all studied $P_{\rm L1}$, the increase of $P_{\rm L2}$ and, in turn, $n_2$ leads to $\Delta E_2$ larger in absolute value than $\Delta E_1$, indicating that the repulsive ${\rm IX}_2 - {\rm IX}_2$ interaction is stronger than the attractive ${\rm IX}_2 - {\rm IX}_1$ interaction.

\begin{figure}
\begin{center}
\includegraphics[width=8.5cm]{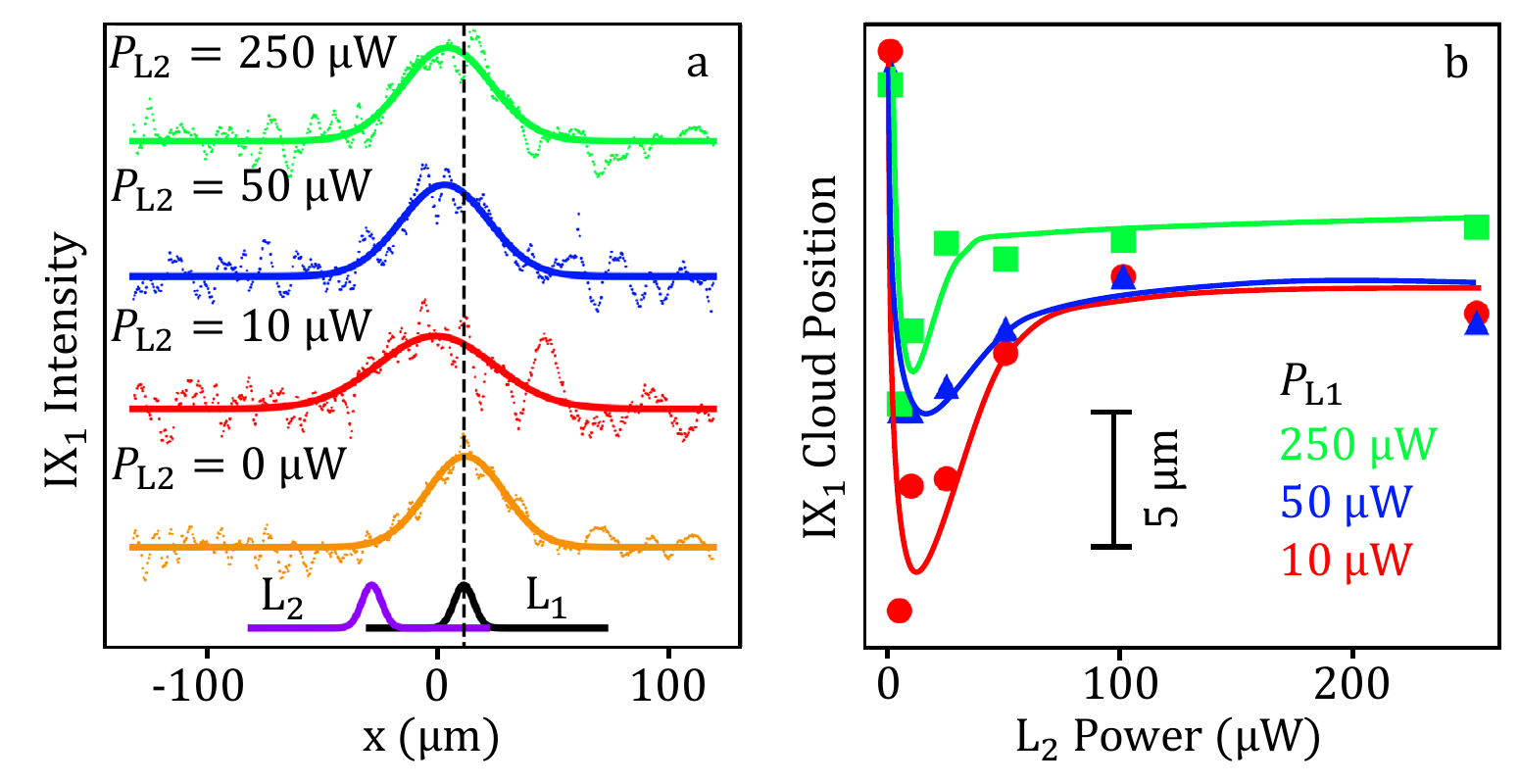}
\caption{\textbf{Attraction of the IX$_1$ cloud to the IX$_2$ cloud: Experiment.} 
(a) The IX$_1$ cloud profiles at different $P_{\rm L2}$ and, in turn, IX$_2$ densities. $P_{\rm L1} = 10$~$\mu$W. The profiles of L$_1$ and L$_2$ laser excitation spots are shown by black and purple lines, respectively. Dashed line indicates the center of L$_1$ excitation spot.
(b) The center of mass position of the IX$_1$ cloud as a function of $P_{\rm L2}$ for different $P_{\rm L1}$.
}
\end{center}
\label{fig:spectra}
\end{figure}

We also consider the spatial shift of the IX$_1$ cloud toward the IX$_2$ cloud~\cite{IX15shift}. Figure~3 shows that the IX$_1$ cloud attracts to the IX$_2$ cloud. With increasing $P_{\rm L2}$, the spatial shift is nonmonotonic. This behaviour is observed for different $P_{\rm L1}$. A larger spatial shift, reaching $\sim 10$~$\mu$m, is observed at low $P_{\rm L1}$ (Fig.~3b). 

\begin{figure*}
\begin{center}
\includegraphics[width=15.5cm]{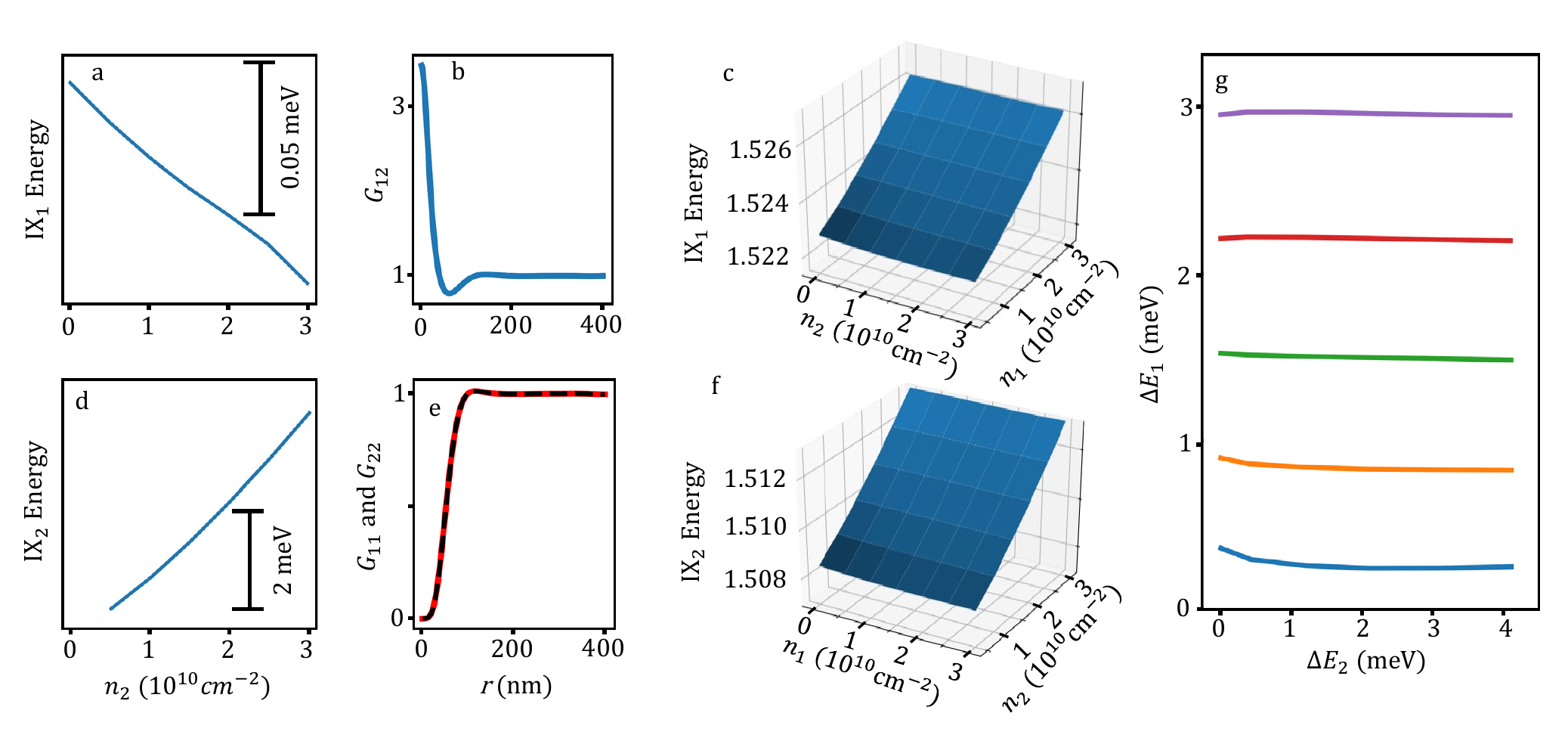}
\caption{\textbf{Decrease and increase of IX energy due to attractive and repulsive dipolar IX interaction: Theory.}
(a) The decrease of IX$_1$ energy with increasing IX$_2$ density. $n_1=1.5 \times 10^{10}$~cm$^{-2}$. An energy decrease corresponds to attractive ${\rm IX}_2 - {\rm IX}_1$ interaction.
(b) ${\rm IX}_1 - {\rm IX}_2$ density correlation function. $n_1 = n_2 = 10^{10}$ cm$^{-2}$. 
(c) IX$_1$ energy as a function of both IX$_1$ and IX$_2$ density.
(d) The increase of IX$_2$ energy with increasing IX$_2$ density. $n_1 = 0$. An energy increase corresponds to repulsive ${\rm IX}_2 - {\rm IX}_2$ interaction. 
(e) ${\rm IX}_2 - {\rm IX}_2$ (red solid line) and ${\rm IX}_1 - {\rm IX}_1$ (black dashed line) density correlation functions. $n_1 = n_2 = 10^{10}$ cm$^{-2}$.
(f) IX$_2$ energy as a function of both IX$_1$ and IX$_2$ density. 
(g) The change in IX$_1$ energy vs the change of IX$_2$ energy. Each line corresponds to increasing IX$_2$ density. For the blue, orange, green, red, and purple lines, $n_{\rm 1} = 0.5, 1, 1.5, 2,$ and $2.5 \times 10^{10}$~cm$^{-2}$, respectively.
}
\end{center}
\label{fig:spectra}
\end{figure*}

{\bf Theory.} We analyze the dipolar interaction in IX bilayers theoretically and compare the experimental data with theoretical simulations. The numerical simulation of such two-species many-body system is done through the Hyper-netted Chain (HNC) formalism~\cite{Chakraborty1982}. The intra- and inter-layer interaction is modelled by assuming that the wave-function of electrons and holes are isotropic Gaussians and they experience a Coulomb interaction, see SI. HNC method has been previously used for studying bosons with dipolar interactions in single-layer systems~\cite{Abedinpour2012}. Unless the interaction strength is very high, the HNC predictions for basic many-body properties such as pair-correlation functions and energy per particle were shown to be in a good agreement with the more accurate Monte-Carlo calculations~\cite{Astrakharchik2007, Macia2014, Cinti2017}.

The simulated IX energy shifts caused by the attractive and repulsive IX dipolar interactions are presented in Fig.~4. Figures~4a,c show that the increase of $n_2$ leads to a monotonic decrease of IX$_1$ energy due to the attractive ${\rm IX}_1 - {\rm IX}_2$ interaction, in qualitative agreement with the experimental data in Fig.~2a,c. In contrast, the increase of $n_2$ (or $n_1$) leads to a monotonic increase of IX$_2$ (or IX$_1$) energy (Fig.~2d,f) due to the repulsive ${\rm IX}_2 - {\rm IX}_2$ (or ${\rm IX}_1 - {\rm IX}_1$) interaction, in qualitative agreement with the experimental data in Fig.~2d,f. 

In the simulations, $n_1$ is increased selectively keeping $n_2$ intact. In the experiment, an increase of $n_1$ is accompanied by an increase of $n_2$ as outlined above. This leads to different variations of IX$_2$ energy with $n_1$ in the experiment (Fig.~2f) and the theory (Fig.~4f). However, the conclusions on the attractive ${\rm IX}_1 - {\rm IX}_2$ interaction and repulsive ${\rm IX}_2 - {\rm IX}_2$ interaction derived from the experiment and the comparison between the experiment and the theory are based on the $n_2$ dependence and are not affected by the difference in the $n_1$ dependence.

The density correlation functions for the cases of attractive ${\rm IX}_1 - {\rm IX}_2$ and repulsive ${\rm IX}_1 - {\rm IX}_1$ and ${\rm IX}_2 - {\rm IX}_2$ interactions are presented in Fig.~4b and 4e, respectively. For the attractively interacting IXs (Fig.~4b), the correlation function enhancement above 1, the mean-field value, increases the interaction energy compared with the vanishing interlayer interaction in the mean-field approximation. On the contrary, the repulsively interacting IXs avoid each other (Fig.~4e) that lowers the intralayer IX interaction energy compared with the uncorrelated state assumed in the mean-field approximation. 

As for the experimental data, we compare the attractive and repulsive dipolar interactions in a graph showing $\Delta E_1$ vs $\Delta E_2$. Figure~4g shows that for all studied $n_1$, the increase of $n_2$ leads to calculated $\Delta E_2$ larger in absolute value than $\Delta E_1$, in qualitative agreement with the experimental data in Fig.~2g. 

We also simulated the spatial shifts of the IX$_1$ cloud toward the IX$_2$ cloud. The simulations of the IX spatial profiles are based on the IX generation, diffusion, and recombination and are outlined in SI.  Figure~5 shows that the IX$_1$ cloud attracts to the IX$_2$ cloud in the simulations, in qualitative agreement with the attraction observed in the experiment (Fig.~3). In comparison, both our experimental measurements (Fig.~S3) and theoretical simulations (Fig.~S4) show that two clouds of repulsively interacting IX$_2$ repel each other, see SI.  

\begin{figure}
\begin{center}
\includegraphics[width=8.5cm]{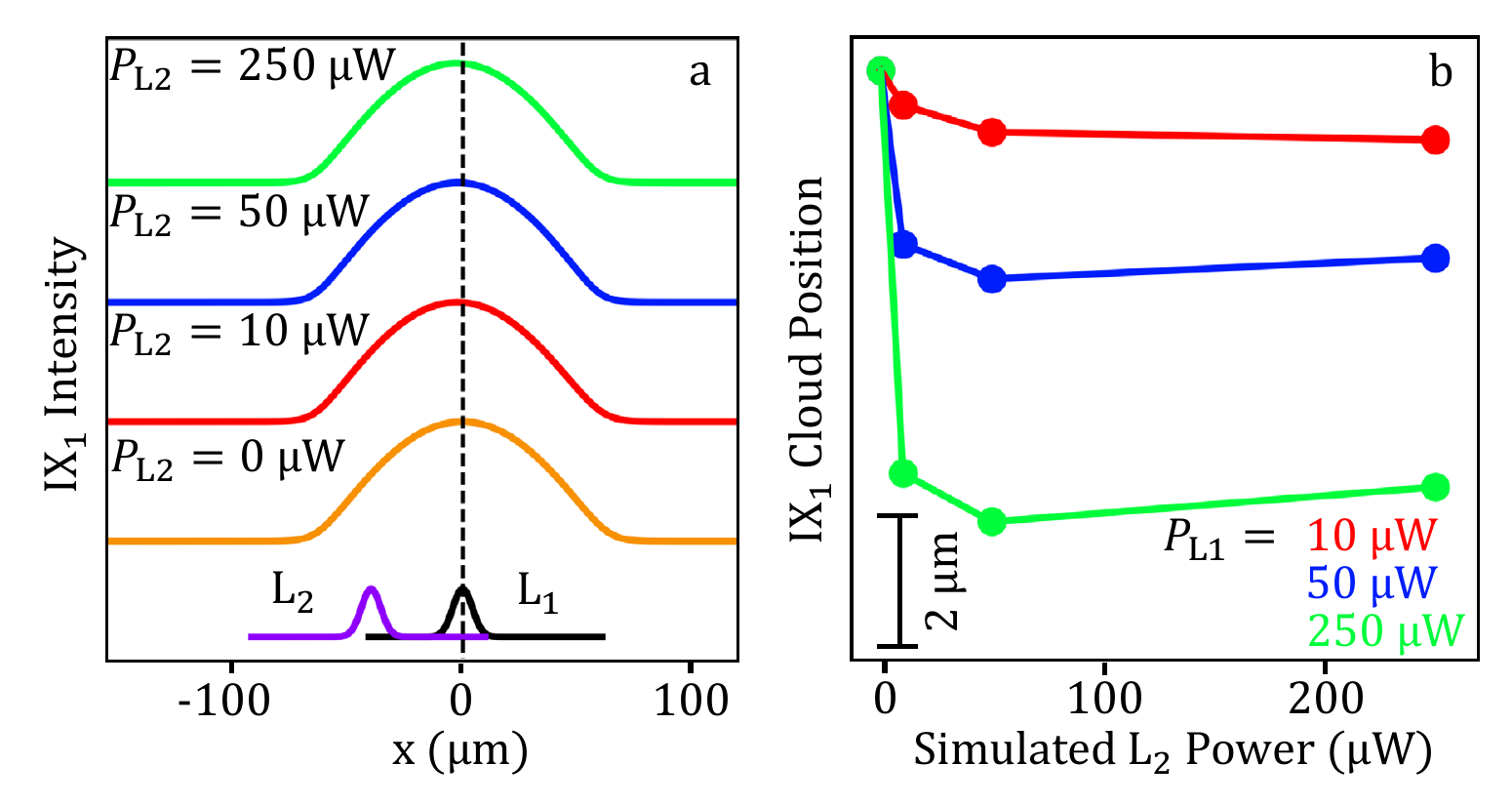}
\caption{\textbf{Attraction of the IX$_1$ cloud to the IX$_2$ cloud: Theory.}
(a) The IX$_1$ cloud profiles simulated for different $P_{\rm L2}$ and, in turn, IX$_2$ densities.
$P_{\rm L1} = 250$~$\mu$W. 
The profiles of L$_1$ and L$_2$ laser excitation spots are shown by black and purple lines, respectively. Dashed line indicates the center of L$_1$ excitation spot.
(b) The center of mass position of the IX$_1$ cloud as a function of $P_{\rm L2}$ for different $P_{\rm L1}$.
}
\end{center}
\label{fig:spectra}
\end{figure}

While both the experimental measurements and theoretical simulation show (i) a monotonic IX$_1$ energy reduction with increasing IX$_2$ density and (ii) an in-plane shift of IX$_1$ cloud towards IX$_2$ cloud, consistent with attractive dipolar interaction, the measured IX$_1$ energy reduction and IX$_1$ cloud shift are higher than the values given by the correlated liquid theory (compare Figs.~2a and 4a, Figs.~2g and 4g, and Figs.~3 and 5). 

The interaction- and/or disorder-induced IX mass enhancement may be one possible reason for this discrepancy. The magnitude of IX$_1$ energy reduction, $\Delta E_1$, scales with the strength of  interlayer ${\rm IX}_1 - {\rm IX}_2$ dipolar attraction (Fig.~S1). For the case of a single ${\rm IX}_1 - {\rm IX}_2$ pair, $\Delta E_1$ can be estimated as the binding energy of ${\rm IX}_1 - {\rm IX}_2$ biexciton state, $E_b$. For a bare IX mass, $m_{\rm IX} \sim 0.2 m_0$~\cite{Butov2001}, $E_b \sim 0.3$~meV (Fig.~S2). Higher $E_b$ can be achieved for higher IX masses, and, e.g. for the reduced IX mass enhanced to $2 m_0$, $E_b$ reaches $\sim 1.2$~meV (Fig.~S2), making the ${\rm IX}_1 - {\rm IX}_2$ interaction scale comparable to the experiment (Fig.~2). A mass enhancement can be caused by interaction, however, only a relatively weak interaction-induced mass enhancement, up to $\sim 25 \%$, was observed in electron-hole systems in single QWs~\cite{Butov1992}. The studies of effects of interaction and/or disorder on the IX mass can be a subject for future work. 

In summary, we presented experimental and theoretical studies of attractive dipolar interaction in IX bilayers. We found that increasing density of IXs in one layer causes a monotonic energy reduction for IXs in the other layer. We also found an in-plane shift of a cloud of IXs in one layer towards a cloud of IXs in the other layer. This behaviour is qualitatively consistent with attractive dipolar interaction. The measured IX energy reduction and IX cloud shift are higher than the values given by the correlated liquid theory.

These studies were supported by DOE Office of Basic Energy Sciences under award DE-FG02-07ER46449. The heterostructure growth and fabrication were supported by SRC and NSF grant 1905478. The theoretical studies were supported by Office of Naval Research grant ONR-N000014-18-1-2722.


\begin{references}
\bibitem{3D}
The dipolar interaction $v(r) \propto 1/r^3$ is long-range for 3D and short-range for 2D and 1D systems.
\bibitem{IX15shift}
Due to the longer IX$_{2}$ lifetime, the IX$_{2}$ PL signal is lower than the IX$_{1}$ PL signal and the IX$_{2}$ cloud is wider than the IX$_{1}$ cloud. These factors make the accuracy of determining the IX$_{2}$ cloud shift too low to analyze this shift.
\end{references}
\end{document}


\date{\today}

\title{Supporting information: Attractive and repulsive dipole-dipole interaction between indirect excitons}

\author{D.\,J.~Choksy}
\affiliation{Department of Physics, University of California at San Diego, La Jolla, California 92093-0319, USA}
\author{Chao~Xu}
\affiliation{Department of Physics, University of California at San Diego, La Jolla, California 92093-0319, USA}
\author{M.\,M.~Fogler}
\affiliation{Department of Physics, University of California at San Diego, La Jolla, California 92093-0319, USA}
\author{L.\,V.~Butov}
\affiliation{Department of Physics, University of California at San Diego, La Jolla, California 92093-0319, USA}
\author{J.~Norman}
\affiliation{Materials Department, University of California at Santa Barbara, Santa Barbara, California 93106-5050, USA}
\author{A.\,C.~Gossard}
\affiliation{Materials Department, University of California at Santa Barbara, Santa Barbara, California 93106-5050, USA}

\date{\today}

\begin{abstract}
\noindent

\end{abstract}
\maketitle

\renewcommand*{\thefigure}{S\arabic{figure}}

\section{Exciton interaction}
\subsection{Interaction potentials}

\begin{figure}
	\center
	\includegraphics[width=3.0in]{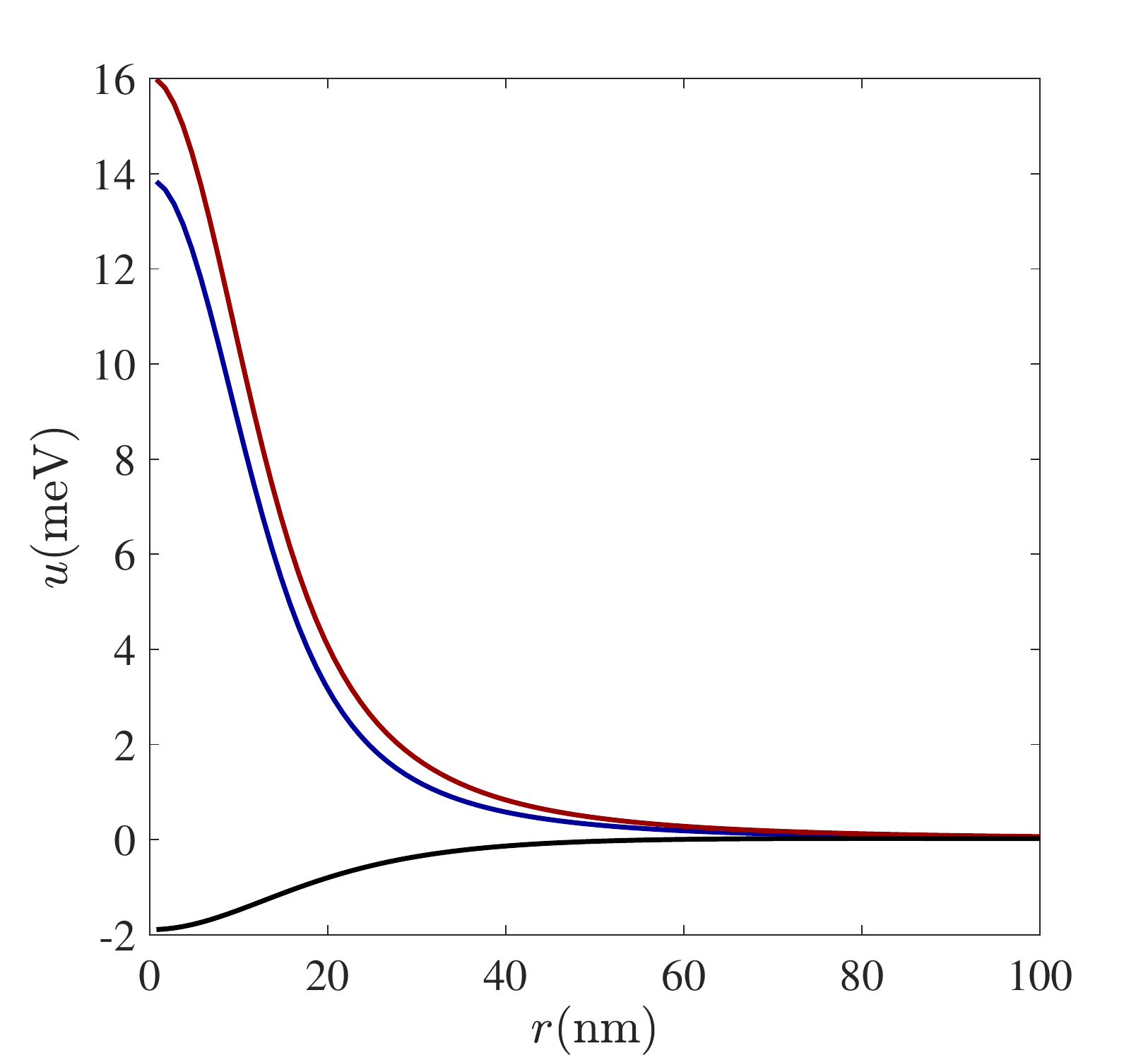}
	\caption{Model interaction potentials: $u_{22}$ (top) and $u_{11}$ (middle) are the intra-CQW potentials for ${\rm{IX}}_2$ and ${\rm{IX}}_1$, respectively; $u_{12}$ (bottom curve) is the inter-CQW interaction potential.
	Parameters: $d_1 = 20$, $d_2 = 25$, $D = 43$, $a = 5$ (all in nm); $\epsilon = 13$.
	}
	\label{fig:potentials}
\end{figure}

We modeled {\rm{IX}}s as composite bosonic particles with
a rigid internal charge distribution.
The interactions of such particles can be specified in terms of potentials $u_{ij}(r)$, $1 \leq i, j \leq 2$, which are functions of
pairwise in-plane distances $r$ of the excitons.
Here and below we use subscripts $1$ and $2$ to label the CQWs ($12\,\text{nm}$- and $15\,\text{nm}$-wide, respectively).
To compute these interaction potentials, we assumed that the charge distributions of all the electrons and holes are spherically symmetric Gaussians.
The radius $a$ of the Gaussians is our adjustable parameter that accounts for the
width of the quantum wells and the internal motion of particles about the center of mass of each exciton.
We computed these interaction potentials by taking the convolutions of the Coulomb kernel $e^2 / \epsilon \sqrt{r^2 + z^2}$
with the charge densities of the interacting particle pairs.
The result for the inter-CQW potential $u_{12}(r) = u_{21}(r)$ is
\begin{align}
    u_{12}(r) &= \sum_{\sigma = \pm} \sum_{\tau = \pm}
    V\left(\sqrt{r^2 + z_{\sigma \tau}^2}\, \right),\quad
	V(r) = \frac{e^2}{\epsilon r}\, {\mathrm{erf}}\left(\frac{r}{2a}\right),
	\quad
    z_{\sigma \tau} = \tau \frac{d_1 - \sigma d_2}{2} + D\,,
\label{eqn:u_12}
\end{align}
where ${\mathrm{erf}}(x)$ is the error function and
$D$ is the $z$-axis distance between the CQW centers.
The intra-CQW potentials $u_{k k}(r)$, $k = 1$ or $2$, are given by the same equation with
$D = 0$ and the electron-hole separations $d_1$, $d_2$ replaced by $d_k$.
The plots of these potentials for parameter values representative of our experimental device are shown in Fig.~\ref{fig:potentials}.
At $r \ll a$ all of them approach constant finite values and 
at large $r$, these potentials behave as $1 / r^3$.
Potentials $u_{11}(r)$ and $u_{22}(r)$ are strictly repulsive. Potential $u_{12}(r)$ is attractive in the range of distances $r$ selected for the plot. At larger $r$, it eventually becomes repulsive but it is already very small as such $r$.

The potentials $u_{ij}(r)$ serve as inputs to our computer program that computes many-body properties within the zero-temperature hypernetted chain (HNC) formalism.
Another input parameter is the effective mass $m_\mathrm{IX}$ of the excitons, which we took to be $0.2$ of the free electron mass $m_0$.
Our implementation of the HNC method is based on Ref.~\cite{Chakraborty1982}.
The output of these calculations include the pair correlation functions $G_{ij}(r)$,
the energy density $\varepsilon = \varepsilon(n_1, n_2)$, and the chemical potentials
\begin{align}
	\mu_j = \partial \varepsilon / \partial n_j
\label{eqn:mu}
\end{align}
of the excitons as functions of their number densities $n_1$ and $n_2$ in the CQWs.
If the shake-up effects, i.e., many-body relaxation processes following the exciton recombination,
can be neglected,
then the exciton emission energies (or exciton ``single-particle energies'') $E_j$
should coincide with their chemical potentials:
\begin{equation}
E_j \approx \mu_j .
\label{eqn:E_j_mu_j}
\end{equation}
Based on this assumption,
we have constructed the plot of $\Delta E_{1}$ \textit{vs}. $\Delta E_{2}$ shown in Fig.~4g of the main text. Representative intra-CQW and inter-CQW pair-correlation functions are plotted in Fig.~4b,e. At short distances, these functions show a deep ``correlation hole'' for excitons of the same CQW and a strong correlation peak for excitons of different CQWs.

\subsection{Biexciton binding energy}

\begin{figure}
	\center
	\includegraphics[width=3in]{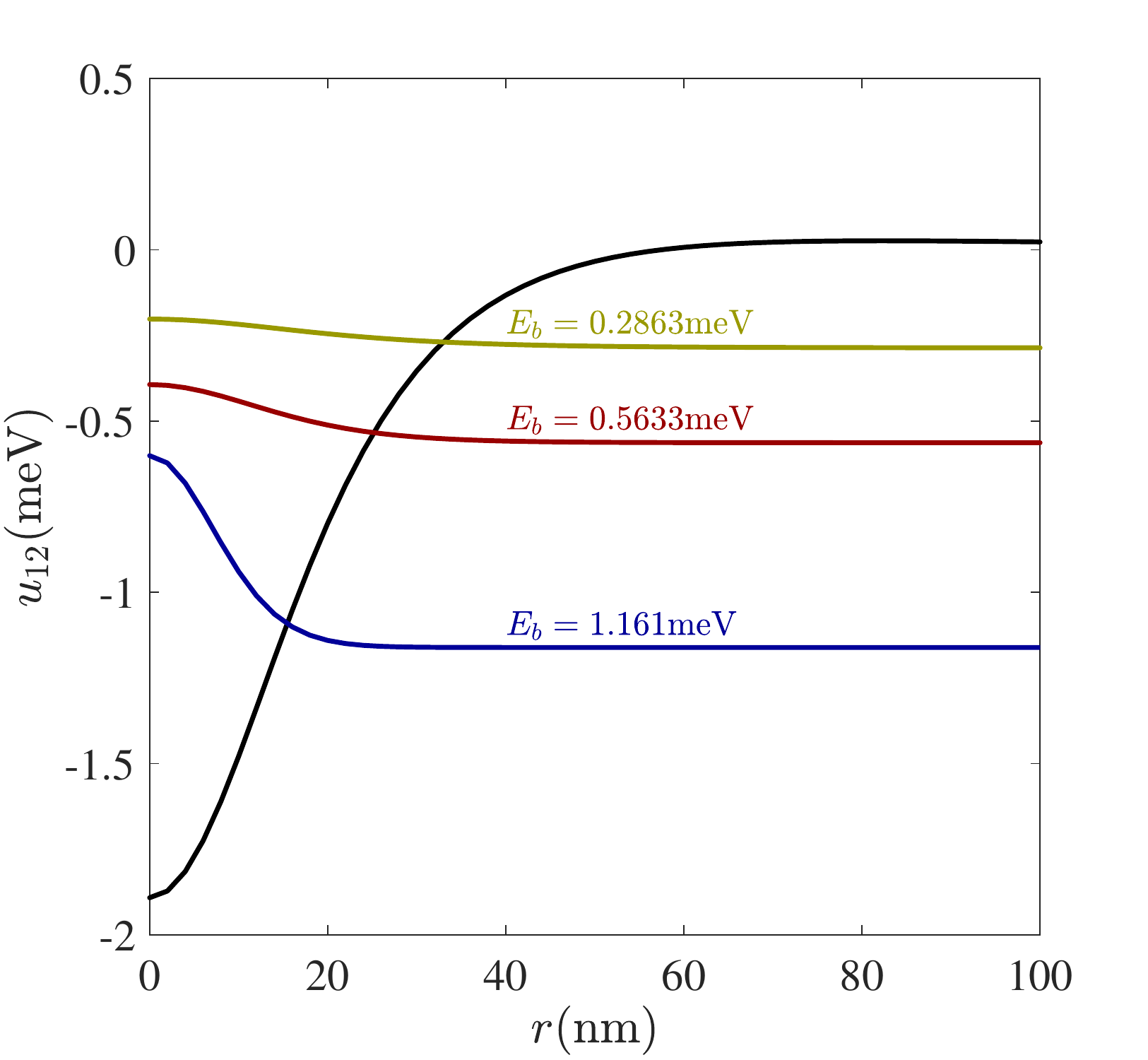}
	\caption{Probability density distributions of inter-CQW biexcitons within the rigid-body approximation.
		The interaction potential $u_{12}$ (same as in Fig.~\ref{fig:potentials}) is shown by the curve with a dip (black curve). The probability distributions (wavefunctions squared) of the bound states are depicted by the curves with the maxima. These curves are plotted in arbitrary units and are offset to show the binding energies $E_b$. They correspond (top to bottom) to
		the reduced masses of $0.5$, $1.0$, and $10$ of the exciton mass.
	}
	\label{fig:bound_state}
\end{figure}

From previous theoretical work on double-layer bosonic systems with repulsive intra-layer and attractive inter-layer dipolar interactions~\cite{Lu2008, Safavi-Naini2013, Macia2014, Cinti2017},
we expect that the exciton system at low enough equal densities $n_1 = n_2$
should be made of bound pairs, the inter-CQW biexcitons.
If $n_1 \neq n_2$, then biexcitons and unpaired excitons may co-exist.
A rough estimate of the required density is given by the Mott criterion
stating that the biexcitons should appear when the dimensionless parameter
$\min (n_1, n_2)\, b^2$ is less than some critical number,
which is usually numerically small, perhaps, $0.02$.
Here $b$ is the spatial size of the biexciton.
Within our rigid-body approximation,
the biexciton bound state can be easily found numerically.
In the relative coordinates, this problem reduces to solving a Schr\"odinger equation for a particle of reduced mass $m_\mathrm{IX} / 2$ 
subject to the confining potential $u_{12}(r)$.
For the same parameters as in Fig.~\ref{fig:potentials}, we obtained the binding energy to be $E_b = 0.286\,\mathrm{meV}$.
From Fig.~\ref{fig:bound_state} we deduce the spatial size of the biexciton to be $b \sim 30\,\mathrm{nm}$, so that the Mott critical density 
for biexcitons is
of the order of $10^{10}\,\mathrm{cm}^{-2}$, not too far from the exciton densities realized in our experiment.

The following argument suggests that $E_b$ is in fact the maximum possible shift of the single-particle energies due to the inter-CQW attraction.
Indeed, in the limit of high densities, where average intra-CQW exciton
separation is smaller than $D$, correlations are negligible.
At intermediate densities, where HNC should be accurate,
the dependence of say $E_1 = E_1(n_1, n_2)$ on $n_2$ with $n_1$ held fixed is either monotonic or flat within computational accuracy, see Fig.~4c,f.
Therefore, the asymptotic limit $n_1 = n_2 \to 0$,
where all excitons are paired and
\begin{equation}
E_1 = E_2 = -E_b
\label{eqn:E_j_as}
\end{equation}
should correspond to the largest possible attraction effect.
Note that the HNC method reproduces this asymptotic limit only
approximately.
The tendency toward pairing is manifested in the aforementioned peak in the
pair-correlation function $G_{12}(r)$ at $r = 0$.
The shape of this peak computed by the HNC resembles the probability distribution of the biexciton, cf.~Figs.~4b 
and \ref{fig:bound_state}.
The integrated weight $N = n_2 \int G_{12} (r)d^2 r$ of the peak
(where the integration extends up to $r \sim b$)
is the average number of excitons of CQW$_1$ attracted to an exciton in CQW$_2$.
When the biexcitons form, $N$ should approach unity.
Yet within our HNC calculations $N$ keeps increasing as $n_1 = n_2$ decreases.
This suggests that the standard HNC method is inadequate in the low-density regime where we should instead use Eq.~\eqref{eqn:E_j_as}.

We found both in experimental measurements (Fig.~2a,c) and theoretical simulations (Fig.~4a,c) that increasing density of IXs in one layer causes a monotonic energy reduction for IXs in the other layer. These results differ with the results of Refs.~\cite{Hubert2019, Hubert2020} where a nonmonotonic dependence on the density was reported.
The nonmonotonic dependence on the density was attributed to many-body polaron effects in Refs.~\cite{Hubert2019, Hubert2020}. Our simulations show no indication for the non-monotonic dependence on the density.

The experiment still poses a challenge for the theory
because the shift of $E_1$ has been observed to routinely exceed the computed $E_b = 0.286\,\mathrm{meV}$, see Fig.~2a.
To identify a possible reason for the discrepancy, we examined this important parameter more critically.
First, we tested the validity of the rigid-body approximation.
We used a previously developed computational tool~\cite{Meyertholen2008}
to accurately solve for the exciton and biexciton binding energies as two-body and four-body problems, respectively.
For the parameters of Fig.~\ref{fig:potentials} we obtained
$E_b = 0.33\,\mathrm{meV}$.
Hence,
the rigid-body approximation is not the major source of the discrepancy.
Next, we noticed that $E_b$ is greatly reduced compared to 
the depth $\approx 2\,\mathrm{meV}$ of the potential well $u_{12}(r)$.
This reduction is due to the zero-point motion.
As an illustration of how this quantum effect may affect the binding energy,
we recalculated $E_b$ and the wavefunctions of biexcitons for reduced masses
enhanced two- and twenty-fold.
In the latter case, the binding energy rises to $1.16\,\mathrm{meV}$, see Fig.~\ref{fig:bound_state},
which is close to the experimentally measured shifts
of $E_1$ we attributed to the inter-CQW attraction.
It is hard to expect that the exciton mass is indeed enhanced by such an enormous factor due to the interaction alone. (As a point of reference, only a relatively weak interaction-induced mass enhancement, up to $25 \%$, was observed in electron-hole systems in single QWs~\cite{Butov1992}.)
However, the suppression of the zero-point motion of an exciton pair may in principle 
be facilitated by
disorder in the system that traps the excitons close together in deep potential wells.

\section{Dynamics of exciton density distribution}

\begin{figure}
	\begin{center}
		\includegraphics[width=12cm]{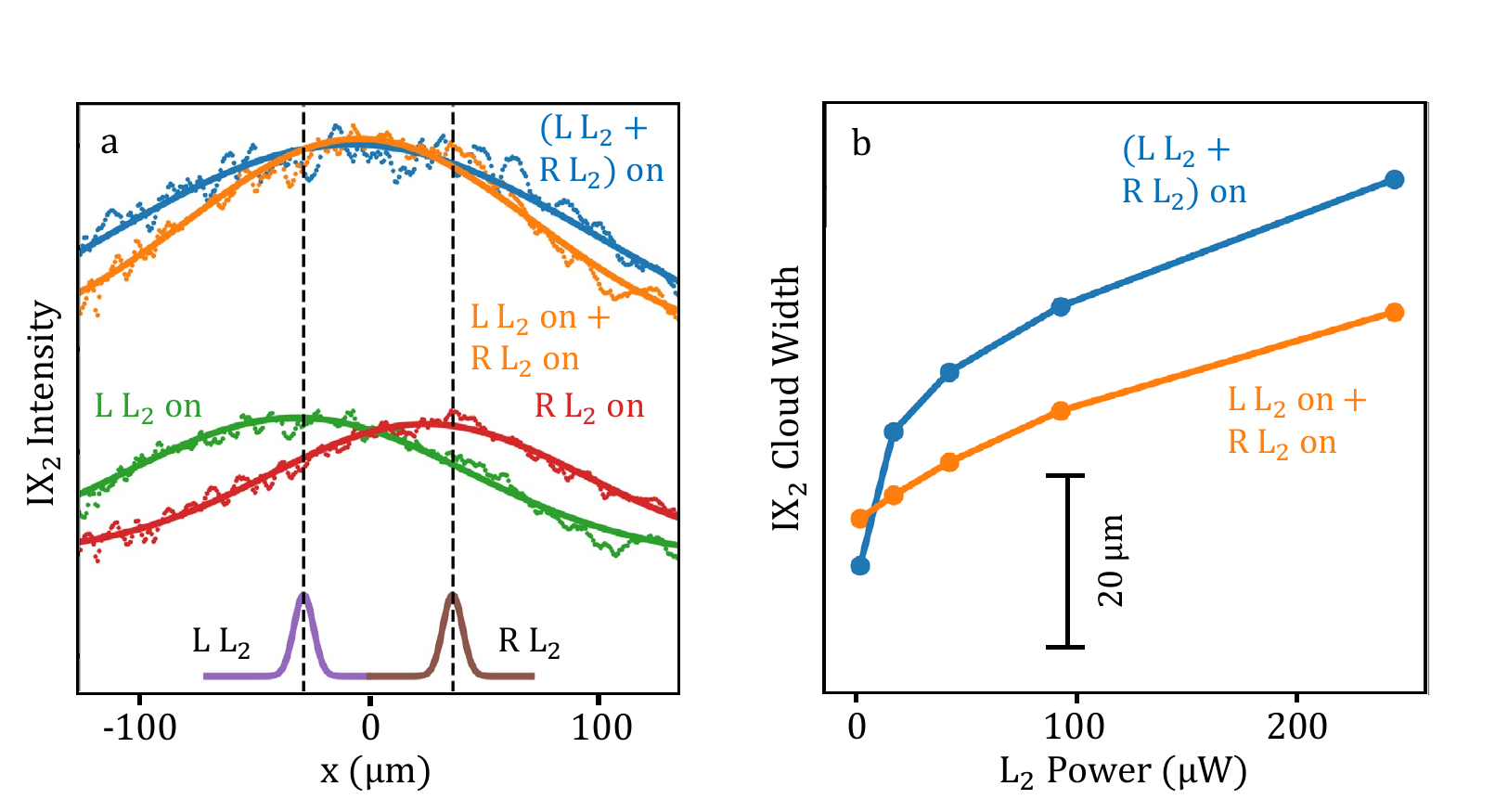}
		\caption{\textbf{Repulsion between the IX$_2$ clouds: Experiment.} 
			(a) The IX$_2$ cloud profiles when only left L$_2$ is on (L L$_2$ on, green line), when only right L$_2$ is on (R L$_2$ on, red line), and when both left and right L$_2$ are on [(L L$_2$ + R L$_2$) on, blue line]. The sum of L L$_2$ on profile and R L$_2$ on profile (L L$_2$ on $+$ R L$_2$ on) is shown by orange line. Profile (L L$_2$ + R L$_2$) on is wider than the sum of L L$_2$ on profile and R L$_2$ on profile, indicating the repulsion between the IX$_2$ clouds. The profiles of left and right L$_2$ laser excitation spots are shown by purple and brown lines, respectively. Dashed lines indicate the centers of the excitation spots.
			(b) The width of IX$_2$ cloud when both left and right L$_2$ are on [(L L$_2$ + R L$_2$) on, blue points] in comparison to the width of IX$_2$ cloud obtained as the sum of L L$_2$ on cloud and R L$_2$ on cloud [L L$_2$ on $+$ R L$_2$ on, orange points] as a function of $P_{\rm L2}$.
		}
	\end{center}
	\label{fig:spectra}
\end{figure}

\begin{figure}
	\begin{center}
		\includegraphics[width=12cm]{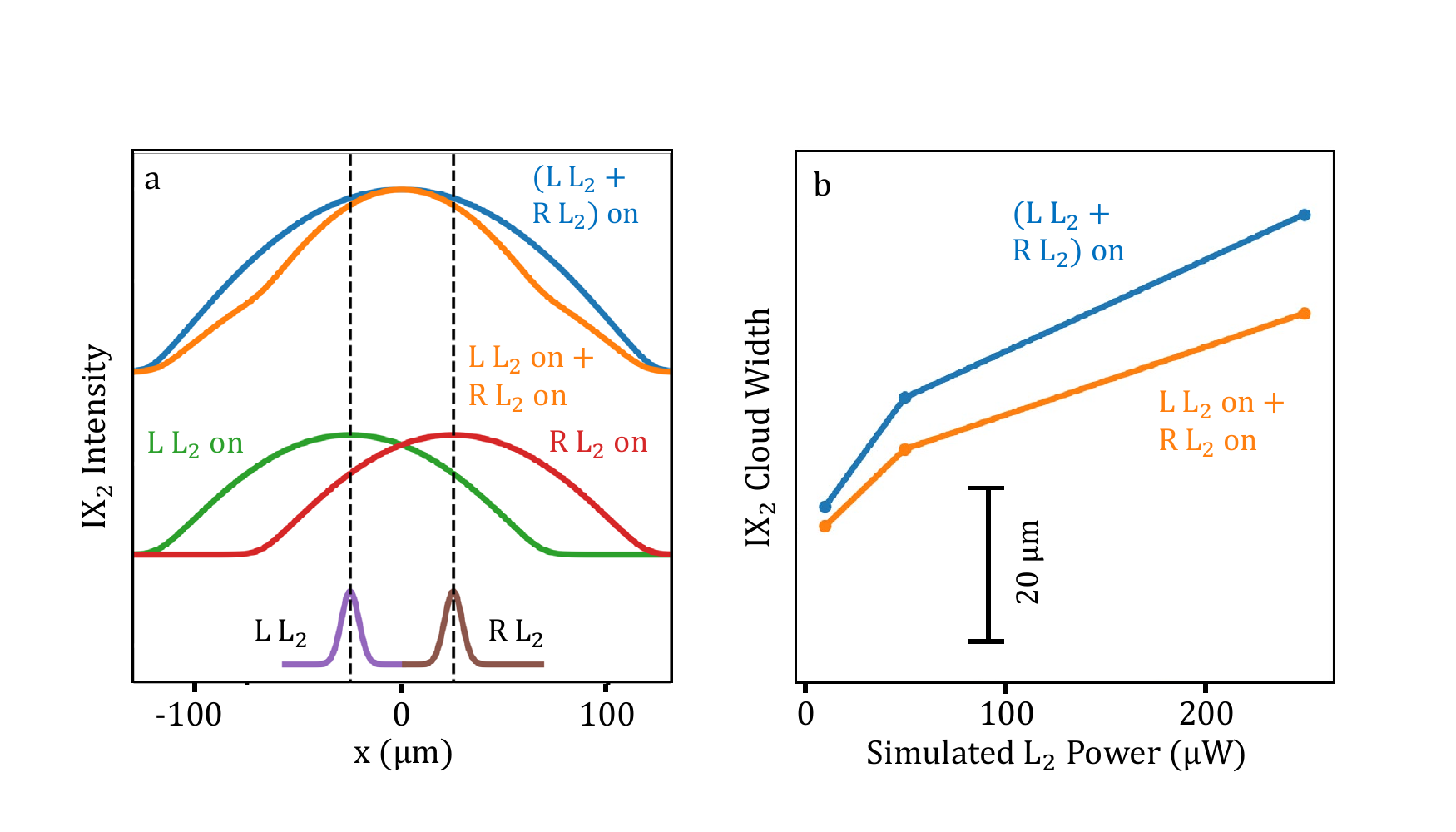}
		\caption{\textbf{Repulsion between the IX$_2$ clouds: Theory.}
			(a) The IX$_2$ cloud profiles simulated when only left L$_2$ is on (L L$_2$ on, green line), when only right L$_2$ is on (R L$_2$ on, red line), and when both left and right L$_2$ are on [(L L$_2$ + R L$_2$) on, blue line]. The sum of simulated L L$_2$ on profile and R L$_2$ on profile (L L$_2$ on $+$ R L$_2$ on) is shown by orange line. Profile (L L$_2$ + R L$_2$) on is wider than the sum of L L$_2$ on profile and R L$_2$ on profile, indicating the repulsion between the IX$_2$ clouds. The profiles of left and right L$_2$ laser excitation spots are shown by purple and brown lines, respectively. Dashed lines indicate the centers of the excitation spots.
			(b) The width of simulated IX$_2$ cloud when both left and right L$_2$ are on [(L L$_2$ + R L$_2$) on, blue points] in comparison to the width of IX$_2$ cloud obtained as the sum of simulated L L$_2$ on cloud and R L$_2$ on cloud [L L$_2$ on $+$ R L$_2$ on, orange points] as a function of $P_{\rm L2}$.
		}
	\end{center}
	\label{fig:spectra}
\end{figure}

In this section we summarize the set of equations we used to model the macroscopic dynamics of excitons.
To simplify the modeling, we assumed that the exciton densities $n_{k}$ and currents $j_{k}$ were functions of a single spatial coordinate $x$.
These quantities obey the continuity equation
\begin{align}
	\partial_t n_{k}(x,t) &= \partial_x j_{k} + g_{k}(x,t) - n_{k}/\tau_{k}\,,
\end{align}
where $\tau_{k}$ is the lifetime of the excitons in $k$th CQW, which is known from the experiment,
and $g_{k}$ is the generation rate proportional to the local laser power.
To represent the exciton currents, we used the drift-diffusion approximation,
\begin{align}
	j_{k}(x,t) &= -D_{k}\partial_x n_{k}(x,t) - B_{k} n_{k}(x,t) \partial_x \mu_{k}\,,
\end{align}
where $D_k$ and $B_k = D_k / T$ are the diffusion coefficient and the drift mobility, respectively.
Finally, to simplify the treatment of interaction effects, we linearized the
density dependence of the exciton chemical potentials, such that
\begin{align}
    \mu_{1} = \gamma_{11} n_{1} + \gamma_{12} n_{2},
\qquad
    \mu_{2} = \gamma_{22} n_{2} + \gamma_{12} n_{1},
\end{align}
where $\gamma_{ij}$ are interacting constants.
Based on the simulations presented in Figs.~4a and 4d of the main text, we set the constants to be $\gamma_{11} = 9.3$, $\gamma_{22} = 11$, and $\gamma_{12} = -0.2$, 
all in units of $10^{-11}\,\mathrm{meV}\cdot \mathrm{cm}^{-2}$.
[Note that the intra-CQW coupling constants are $1/3$ of the ``plate capacitor'' values, i.e., $\gamma_{kk} = (1/3) \times (4\pi e^2 d_k / \epsilon)$].
We developed a computer program that solves these coupled equations on a discrete grid of $x$ as a function of the time variable $t$, starting from initial conditions $n_1 = n_2 \equiv 0$.
To get a relation between the laser powers and the generation rates, we fitted the shifts $E_k \approx \mu_k$ of the exciton emission energies measured as functions of the laser power to the results of these simulations.
We estimated the diffusion coefficients $D_k$ by fitting the calculated width of the IX$_1$ and IX$_2$  exciton clouds to the measured widths of these clouds generated selectively by L$_1$ or L$_2$.

As outlined in the main text, the simulations show that the IX$_1$ cloud attracts to the IX$_2$ cloud (Fig.~5), in qualitative agreement with the attraction observed in the experiment (Fig.~3). In comparison, both our experimental measurements (Fig.~S3) and theoretical simulations (Fig.~S4) show that two clouds of repulsively interacting IX$_2$ repel each other.

\section{Position-energy luminescence images}

The differential $x$-energy image (Fig.~1b) is obtained by subtracting the $x$-energy images created by only L$_1$ on (Fig.~S5c) and by only L$_2$ on (Fig.~S5b) from the $x$-energy image created by both lasers on simultaneously (Fig.~S5a). 

\begin{figure}
	\begin{center}
		\includegraphics[width=17cm]{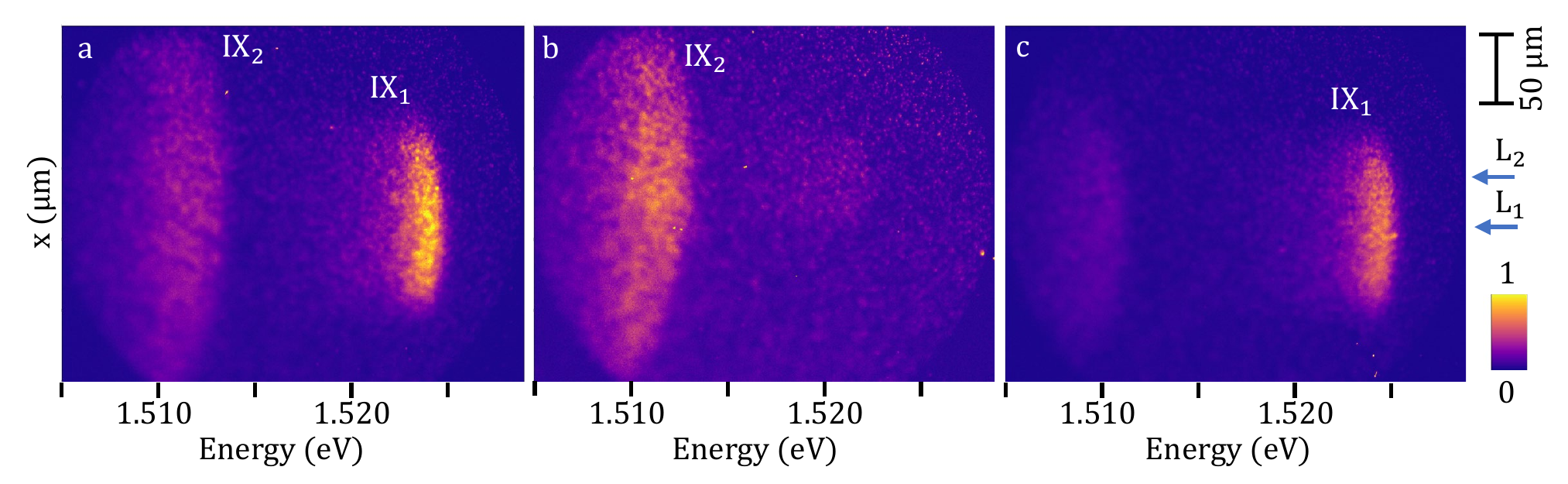}
		\caption{\textbf{Position-energy images of IX luminescence.} 
			(a) Both L$_2$ and L$_1$ lasers are on. (b) Only L$_2$ is on. (c) Only L$_1$ is on. The arrows indicate the excitation spot positions of L$_2$ and L$_1$ lasers resonant to direct excitons in $15\,\text{nm}$ CQW and $12\,\text{nm}$ CQW, respectively. L$_2$ generates IX$_2$. L$_1$ generates IX$_1$ and also a smaller concentration of IX$_2$. The laser powers $P_\mathrm{L1} = 10\,\mu\mathrm{W}$, $P_\mathrm{L2} = 250\,\mu\mathrm{W}$. Gate voltage $V_\mathrm{g} = -2.0\,\mathrm{V}$, temperature $T = 1.7\,\mathrm{K}$.
		}
	\end{center}
	\label{fig:spectra}
\end{figure}